\documentclass[aps,prl,twocolumn,showpacs]{revtex4-1}

\usepackage{graphicx}
\usepackage[ansinew]{inputenc}
\usepackage{array}
\usepackage{color}
\usepackage{amsmath}
\usepackage{amsxtra}
\usepackage{amstext}
\usepackage{amssymb}
\usepackage{latexsym}
\usepackage{float}
\newcommand {\UA}[1]{ {\textcolor{black}{#1}}}

\newcommand {\IR}{{\textsc{IR}}}
\newcommand {\WIR}{{\omega_\IR}}

\begin{document} 

\title{Autoionizing Polaritons in Attosecond Atomic Ionization}

\author{N. Harkema$^1$, C. Cariker$^2$, E. Lindroth$^3$, L. Argenti$^{2,4}$, A. Sandhu$^{1,5}$}
\affiliation{$^1$Department of Physics, University of Arizona, Tucson, Arizona 85721, USA}
\affiliation{$^2$Department of Physics, University of Central Florida, Orlando, Florida 32816, USA}
\affiliation{$^3$Department of Physics, Stockholm University, Stockholm SE-106 91, Sweden}
\affiliation{$^4$CREOL, University of Central Florida, Orlando, Florida 32816, USA}
\affiliation{$^5$College of Optical Sciences, University of Arizona, Tucson, Arizona 85721, USA}
\date{\today}
\email{asandhu@arizona.edu}


\begin{abstract}
Light-induced states are commonly observed in the photoionization spectra of laser-dressed atoms. The properties of autoionizing polaritons, entangled states of light and Auger resonances, however, are largely unexplored. We employ attosecond transient-absorption spectroscopy to study the evolution of autoionizing states in argon, dressed by a tunable femtosecond laser pulse. The avoided crossings between the $3s^{-1}4p$ and several light-induced states indicates the formation of polariton multiplets. We measure a controllable stabilization of the polaritons against ionization, in excellent agreement with \emph{ab initio} theory. Using an extension of the Jaynes-Cummings model to autoionizing states, we show that this stabilization is due to the destructive interference between the Auger decay and the radiative ionization of the polaritonic components. These results give new insights into the optical control of electronic structure in the continuum, and unlock the door to applications of autoionizing polaritons in poly-electronic systems.
\end{abstract}


\maketitle

%

Attosecond pump-probe schemes have opened the way to the real-time control of fast reactions in atoms and molecules~\cite{Johnsson2007, Prince2016}. 
The ionization processes triggered by attosecond pulses, however, limit this control, since they split the target into a parent-ion - photoelectron pair, neither of which is in a pure state~\cite{Goulielmakis2010,Kurka2009,Carlstrom2018}. 
As ionization proceeds, therefore, the coherence of the initial state deteriorates and so does the chance of controlling the subsequent evolution of the products. Ideally, therefore, control should take place while the ionization fragments are still interacting. Autoionizing states (AIS) are natural stepping stones to achieve this goal thanks to their correlated character, high polarizability, and  lifetimes comparable to the duration of an attosecond-pulse sequence.

An AIS differs from a bound state in many respects. In one-photon transitions, its excitation and decay interferes with direct ionization, giving rise to an asymmetric peak in the cross section known as Fano profile~\cite{Fano,Fano_1965}.
%
Pump-probe schemes involving attosecond extreme ultraviolet (XUV) and tunable infrared (IR) dressing pulses, such as transient absorption spectroscopy ~\cite{wu2016review}, have been employed to investigate the dynamics of AIS, demonstrating control not only of their position~\cite{LorentzMeetsFano,He_ATAS_2016,Xe_autoionizing_lifetimes,Ott_Nature2014}, but also of their lifetime, branching ratios, and the XUV transparencies caused by interference with the continuum~\cite{Ott_Nature2014,Loh_He_ATAS_2008,Wang_Ar_ATAS}.

\begin{figure}[hbtp!]
	\includegraphics[width=8cm]{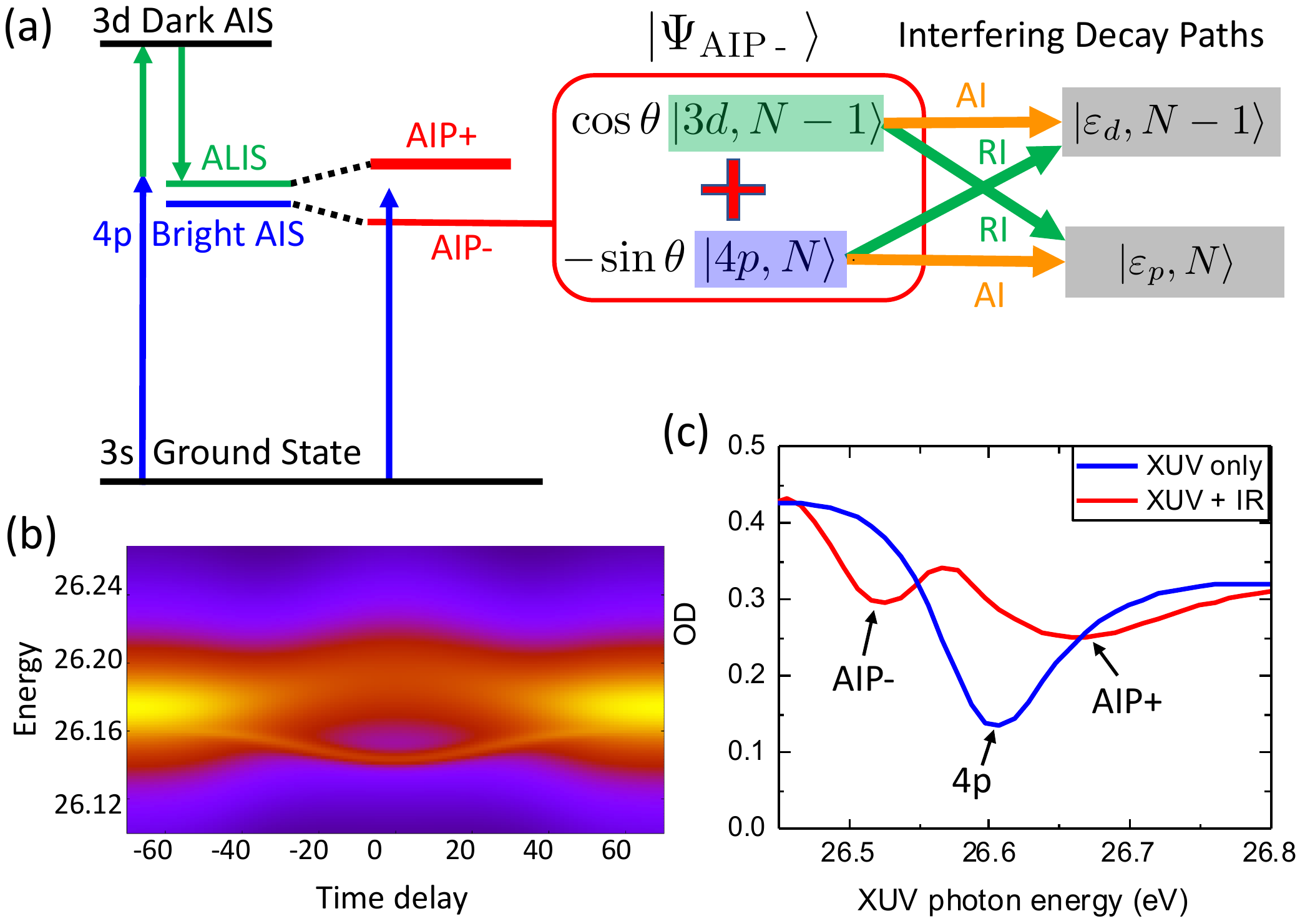}
	\caption{\label{fig:exp_setup} (a) Formation of autoionizing polaritons, AIP$\pm$ from mixing between the bright autoionizing state (AIS) and the autoionizing light-induced state (ALIS) of the dark level. Auger (AI) and radiative (RI) decay paths for the AIPs interfere, leading to stabilization seen as narrow spectral width of the lower branch in simulation (b) and experiment (c).}
\end{figure}

Laser dressed AIS have been studied theoretically since the 1980s ~\cite{Zoller.PhysRevA.24.379,Lambropoulos.PhysRevLett.49.1698,Bachau.PhysRevA.34.4785}, with recent efforts extending to ultrafast XUV spectroscopy ~\cite{Lin.PhysRevA.85.013409,Greene.PhysRevA.85.013411,Argenti2015,Ott_Nature2014}. Since the beginning~\cite{Zoller.PhysRevA.24.379}, theory has indicated the potential of dressing light pulses to stabilize AIS against ionization, as a result of the destructive interference between autoionization and radiative-ionization pathways. This quintessential aspect of electron dynamics in the continuum, which has no counterpart in bound states, has not yet received experimental confirmation. Furthermore, due to the sensitivity of AIS spectral profiles to laser parameters, the quantitative confirmation of their stabilization requires the support of demanding \emph{ab initio} simulations.

In this letter, we report the study of multiple radiatively coupled autoionizing light-induced states (ALIS) in argon, as well as evidence of their stabilization, thus confirming a prediction made nearly forty years ago~\cite{Zoller.PhysRevA.24.379}. We conducted XUV absorption measurements in the presence of a tunable IR pulse, and interpreted them with \emph{ab initio} simulations together with a novel extension of the \UA{Jaynes-Cummings (J-C) analytical model to AIS}. Our results demonstrate quantum control of the AIP decay rate and offer new insights into the properties of ALIS, which are essential for the control of excited electrons prior to ionization in complex poly-electronic systems.

We start with an overview of the extension of J-C model we developed to explain the new phenomenology reported in this work. 
Let $|g\rangle\otimes|N\rangle$ be the ground state of the target atom with $N$ photons in the IR field, and $|\beta\rangle$ a dark AIS. The sequence of states $|\beta\rangle\otimes|N+n\rangle$, with $n=\pm1,\,\pm3,\,\ldots$, are ALIS which appear as a sequence of new autoionizing satellites in the XUV spectrum from the ground state, located symmetrically above ($n>0$) and below ($n<0$) the original dark resonance, and separated from each other by twice the frequency $\WIR$ of the dressing field. 
If the ALIS $|\beta\rangle \otimes |N-1\rangle$ comes close in energy to a bright AIS $|\alpha\rangle \otimes |N\rangle$ with the same parity, the two states split into a pair of autoionizing polaritons (AIP), i.e., entangled states of matter and light.  Fig.\ref{fig:exp_setup}a shows the formation of AIP states $|\Psi_{\textsc{AIP}\pm}\rangle$ that result from the interaction between the bright $3s^{-1}4p$ AIS in argon and the $n=-1$ ALIS of the dark $3s^{-1}3d$ AIS. For this case, the AIP states can be described by the well-known J-C model~\cite{Jaynes1963,Greentree2013}, 
\begin{equation}
\begin{split}
|\Psi_{\textsc{AIP+}}\rangle&=
\cos\theta\, |\alpha\rangle\otimes |N\rangle + \sin\theta\,\, |\beta\rangle\otimes |N-1\rangle,\\
|\Psi_{\textsc{AIP\,-\,\,}}\rangle&=
\sin\theta\,\, |\alpha\rangle\otimes |N\rangle - \cos\theta\, |\beta\rangle\otimes |N-1\rangle.
\end{split}
\end{equation}
The energies of the $|\Psi_{\textsc{AIP}\pm}\rangle$ states are $E_{\pm}=E_\alpha + \delta/2 \pm\Omega/2$, where $\Omega=(\delta^2+\Omega_0^2)^{1/2}$ is the Rabi frequency, $\Omega_0=|\mu_{\alpha\beta}E_0|$, $\mu_{\alpha\beta}=\langle\alpha|\vec{\mu}\cdot\hat{\epsilon}|\beta\rangle$ is the transition dipole moment between the two AIS, $E_0$ and $\hat{\epsilon}$ are the electric field amplitude and polarization, $\delta=E_\beta-E_\alpha-\WIR$ is the detuning, and \textcolor{black}{$\theta=\arctan[\Omega_0/(\Omega-\delta)]$}.

AIPs have unique properties compared to bound-state polaritons, because their decay due to Auger ionization (AI) and the radiative ionization (RI) of their components to the electronic continuum $|\varepsilon\rangle$ can interfere constructively or destructively, depending on the laser coupling parameters, thus enhancing or suppressing their decay rate. Specifically, for the $n=-1$ case above, the two AI and RI decay pathways to the $(N)$-photon sector of the electronic continuum are, respectively,
\begin{equation}
|\alpha\rangle \otimes |N\rangle \longrightarrow |\varepsilon\rangle\otimes |N\rangle,\quad
|\beta\rangle\otimes |N-1\rangle \longrightarrow |\varepsilon\rangle\otimes |N\rangle,
\end{equation}
corresponding to a partial decay rate of the AIP- state \textcolor{black}{
\begin{equation}
\Gamma^-_{N-1} = |\, \sin\theta\,\cdot\, \Gamma_{\textsc{AI},\alpha}^{1/2}\, -\, \cos\theta\,\cdot\, \Gamma_{\textsc{RI},\beta}^{1/2}\,|^2
\end{equation}
}where $\Gamma_{\textsc{AI},\alpha}$ is the field-free AI decay rate of the $\alpha$ AIS, and $\Gamma_{\textsc{RI},\beta} = \pi |\mu_{\varepsilon,\beta}E_0|^2/2$ is the RI rate of the $\beta$ AIS. 
By controlling the field strength and detuning, the amplitudes of the two decay paths can cancel one another, resulting in the stabilization of the AIP. This phenomenon, which we observe in both the experimental measurements and in the \emph{ab initio} simulations, can be regarded as a light-induced transparency in electron-ion scattering, i.e., the matter counterpart of traditional light-induced transparency in atom-XUV-photon scattering~\cite{Fleischhauer.PhysRevLett.84.5094}. 

\UA{The AIP description goes beyond reproducing the well-known Autler-Townes (A-T) splitting, which has been studied extensively by us and others~\cite{Chini2014, Pfeiffer2012, Chini2013_LIS, Wu2013, Loh_He_ATAS_2008, Xe_coreExcitedStates, Harkema2018, Argenti2015}. Specifically, in contrast to the A-T model, it describes the interferences between autoionization and radiative decay of AIS, which is key to the stabilization effect.}


Fig.\ref{fig:exp_setup}b illustrates the model prediction of an AIS pair splitting into two polaritonic states, where the lower branch exhibits a spectral narrowing indicative of its increased lifetime (that is, stabilization) under specific laser intensity and detuning conditions. In these same conditions, when the interference between Auger decay and radiative ionization is not included, the two branches exhibit similar widths, confirming that quantum interference is essential to stabilization. Fig.~\ref{fig:exp_setup}c shows an example of the XUV absorption profile (optical density) of the $3s^{-1}4p$ state in argon, experimentally measured with and without the dressing IR.  Without the IR pulse, we observe the well-known Fano profile of the resonance, with a $q$ parameter of -0.22, in line with the literature~\cite{Madden_Argon_photoabs_1969}. 
When the IR field is present, the spectral profile changes dramatically: the Fano resonance splits into a polaritonic doublet, which manifests itself as two transparency windows with dissimilar widths. The lower branch exhibits a narrow spectral width, indicating stabilization. The properties of the AIP components are highly dependent on the laser detuning, time delay, and intensity. 
In the following, we study this unique phenomenon through the analysis of experimental and theoretical results.
\begin{figure*}[hbtp!]
	\includegraphics[width=18cm]{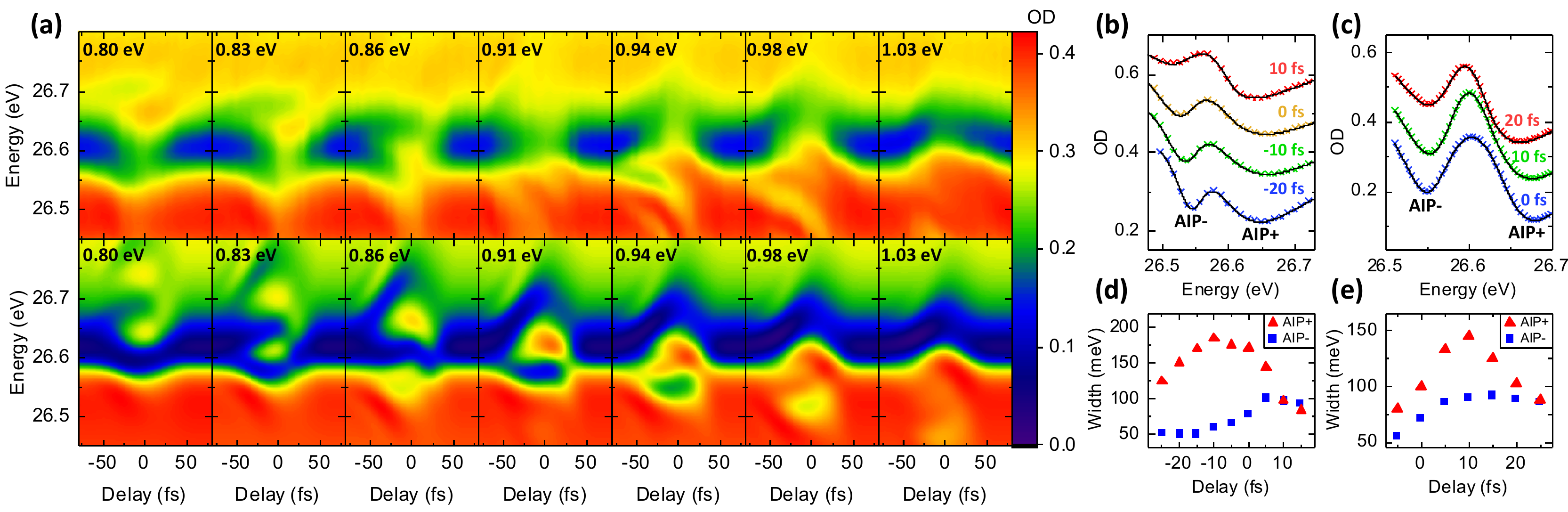}
	\caption{\label{fig:comparison} (a) Experimental (top panels) and theoretical (bottom panels) XUV photoabsorption in the vicinity of argon 3s$^{-1}$4p AIS, as a function of the IR pulse delay, for several values of IR photon energy. Color map represents the OD. Positive delay implies IR pulse arrives before XUV. Interaction of AIS with ALIS gives rise to the polariton splitting prominently visible between 0.86 to 0.98 eV IR photon energy. \textcolor{black}{Experimental (b) and theoretical (c) polaritonic lineshapes for the near resonant case of 0.94 eV IR energy, for several time delays. In absence of interference between radiative and Auger channels, the two polaritons are expected to have comparable widths. Stabilization is evidenced by the delay-dependent reduction of AIP- width and marked difference between the  AIP$\pm$ widths in both experiment (d) and theory (e).}}
\end{figure*}

The experimental setup~\cite{Harkema2018} employs a Ti:sapphire laser amplifier to produce near-infrared (NIR) pulses at 1 kHz with 790 nm central wavelength, 1.8 mJ energy per pulse, and 40 fs pulse duration. A portion (50$\%$) of the beam is focused into a xenon-filled semi-infinite gas cell where it drives high-harmonic generation, producing an XUV attosecond pulse train dominated by the 13th, 15th, and 17th harmonics.  The intensity and wavelength of the driving pulse are adjusted so that the 17th harmonic overlaps with the $3s^{-1}4p$ autoionizing state in argon. The other 50$\%$ of the NIR beam is routed to an optical parametric amplifier, which we use to produce tunable short-wave IR pulses with wavelengths of 1200-1560 nm and durations of 40-90 fs.  

The XUV and tunable IR beams are focused into a 3 mm thick gas cell backed with $\sim$4 torr of argon.  The peak intensity is adjusted to ~40 GW/cm$^2$ at all wavelengths.  The transmitted XUV light is sent to an XUV spectrometer consisting of a reflective concave grating and an x-ray CCD camera.  The IR radiation is filtered out with a 200 nm thick aluminum foil, placed before the CCD.  We measure the \UA{spectrally-resolved intensity of the XUV transmitted through the empty gas cell (no target), $I_0$,} and through the argon-filled gas cell, both in the presence of the IR pulse, $I_\text{XUV+IR}$, and in its absence, $I_\text{XUV}$.  Most XUV transient absorption studies plot the delay-dependent change in optical density due to the IR pulse, $\Delta\text{OD}=-\log(I_\text{XUV+IR}/I_\text{XUV})$. However, in this study, the total optical density $OD = -\log(I_\text{XUV+IR}/I_0)$ is used because it clearly shows how the resonant profiles shift and split as a function of delay. 

The experimental spectra are interpreted with the help of \emph{ab initio} theoretical calculations using the NewStock program~\cite{Carette2013,Marante2017,Chew2018}. The ionic states of the atom are determined, within the electrostatic approximation, by optimizing the energy of the $3s^{-1}$ and $3p^{-1}$ states with a Multi-Configuration Hartree-Fock calculation, performed with the ATSP2K atomic-structure package~\cite{Froese2007}. The full atomic wavefunction is expressed in a close-coupling basis obtained by augmenting the ionic states with a set of radial B-spline and spherical harmonics for the residual electron~\cite{Argenti2016}. To compute the state of the atom $\Psi(t)$ in the presence of the external fields, we solve the time-dependent Schroedinger equation (TDSE) in the close coupling basis, with a unitary split-exponential propagator and within the dipole approximation in velocity gauge. The simulation is accelerated by several orders of magnitude, without affecting the accuracy of the optical observables of interest here, by restricting the basis to a few hundred essential Siegert states~\cite{Siegert1939,Tolstikhin1997,Tolstikhin2006}. The absorption spectrum of the system is computed with the well-known expression for optically thin samples $\sigma(\omega) = \frac{4\pi}{\omega}\Im m\big[\tilde{P}(\omega)/\tilde{A}(\omega)\big]$, \UA{where $P(\omega)$ and $A(\omega)$ are the Fourier transform of the expectation value of the canonical momentum $P(t)=\langle \Psi(t)|\hat{P}_z|\Psi(t)\rangle$ and of the vector potential $A_z(t)$ of the XUV pulse, respectively}. 

Fig.~\ref{fig:comparison} shows the experiment-theory comparison of OD spectrograms for various IR photon energies, exhibiting several recognizable features. At large time delays, the Fano profile of the autoionizing $3s^{-1}4p$ line at 26.6 eV appears as a window resonance, as in the XUV-only lineout in Fig.~\ref{fig:exp_setup}c. Since the autoionization lifetime ($\sim$15 fs) of the $3s^{-1}4p$ state is shorter than the IR pulse duration, it is expected that the IR pulse plays a role only when it overlaps with the XUV pulse. The lineshape of the $3s^{-1}4p$ AIS changes dramatically near zero delay. At all IR photon energies, the depth of the transparency window associated to the $3s^{-1}4p$ state is diminished due to the presence of competing radiative-decay channels.
The $3s^{-1}4p$ lineshape shifts downwards in energy at an IR photon energy of 0.80 and 0.83 eV, and upwards for IR photon energy of 0.94, 0.98, and 1.02 eV.

At the intermediate IR photon energies around 0.9~eV, Fig.~\ref{fig:comparison} reveals that the $3s^{-1}4p$ lineshape splits into two distinct features, which can be identified with AIP$\pm$ from Fig. \ref{fig:exp_setup}a. \UA{In contrast to the two A-T branches of bound states~\cite{Harkema2018,wu2016review}, the two AIP branches appear here as window resonances that can have significantly different widths, at times much narrower than the original $3s^{-1}4p$ resonance. This phenomenon is indicative of the suppressed autoionization rate that results from the destructive interference between the radiative ionization and the Auger decay of the two resonant components of the AIP}. The branch width changes with the time delay and detuning, both of which affect the relative magnitude and phase of the two polariton components, as detailed in the modified J-C model described earlier.

\begin{figure*}[hbtp!]
	\includegraphics[width=15cm]{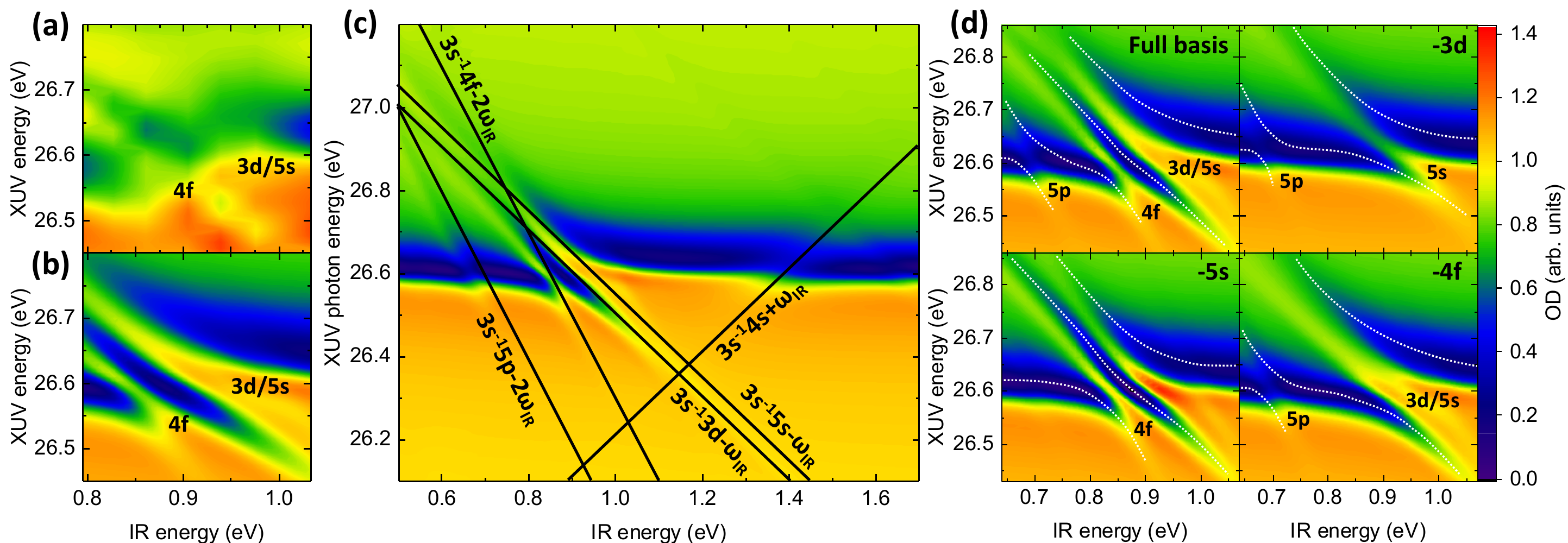}
	\caption{\label{fig:avoidedcrossing} XUV absorption spectra (OD) at $\tau=0$, as a function of the IR photon energy tuning. Experiment (a) and theory (b) show avoided-crossings between the horizontal $4p$ feature and the tilted line of the ALIS from high-lying dark AIS, forming several autoionizing polaritons. (c) Theoretical predictions in a broader spectral and IR tuning range, indicating several other avoided AIS-ALIS and ALIS-ALIS crossings. (d) Theoretical OD as a function of the time delay, with all the essential states involved (top left), or removing, in turn, the $3d$ (top right), $5s$ (bottom left), and $4f$ (bottom right) essential state.}
\end{figure*}
\UA{Figures~\ref{fig:comparison}b,c show the experimental and theoretical XUV absorption lineouts, respectively, at an IR photon energy of 0.94~eV for several time delays. The two plots are in good agreement, showing a broader AIP+ and a narrower AIP- with time-delay dependent widths. We fit these profiles using a formalism that accounts for the decay of two polaritons (AIP${\pm}$) and interference between the Fano amplitudes, such that the absorption cross section is given by 
$\sigma(E) = A + B\left|\frac{\varepsilon_++q_+}{\varepsilon_++i}+C\frac{\varepsilon_-+q_-}{\varepsilon_-+i}\right|^2$
where $\varepsilon_{\pm}=2(E-E_{\pm})/\Gamma_{\pm}$, $E_{\pm}$ are the resonance positions, $\Gamma_{\pm}$ their width, and $q_{\pm}$ are real parameters. Figures~\ref{fig:comparison}d,e show the AIP$\pm$ widths thus obtained, as a function of the time delay, for both experiment and theory. Despite the uncertainties in the experimental delay and IR intensity, experiment and theory are in remarkable agreement. Both show that, at positive delays (IR first), where the IR intensity is small and ionization is dominated by Auger decay, the widths of the polaritonic branches are equal. At progressively lower time delays, on the other hand, as the weight of the competing radiative decay channel increases, the AIP$+$ width grows rapidly, whereas that of AIP$-$ drops, which is evidence of the stabilization of AIP$-$ and the corresponding destabilization of AIP$+$. The maximum value of their ratio is 3 in the experiment and about 1.6 in theory.} \textcolor{black}{At the large negative time delays (IR last), the two branches converge and the line width approaches the static values of $\sim$100 meV in experiment and $\sim$70 meV in theory.}

Apart from the main AIP pair seen at $\sim$0.9 eV, we observe other branches in Fig.~\ref{fig:comparison} as the IR photon energy is varied, indicating the mixing of more than two AIS. We can assign the participating AISs at a given IR photon energy by comparing their known positions with the observed energies $E_i-n\WIR$ of the ALIS states. Fig.~\ref{fig:avoidedcrossing}a-b show the experimental and theoretical absorption spectrum at time delay 0~fs, as a function of the IR-photon energy. Both plots clearly show two avoided crossings: one around $0.85$~eV between the $|3s^{-1}4p\rangle\otimes|N\rangle$ AIS and the $|3s^{-1}4f\rangle\otimes|N-2\rangle$ ALIS, and another around $0.9-0.95$~eV between the $|3s^{-1}4p\rangle\otimes|N\rangle$ AIS and the almost coincident $|3s^{-1}3d/5s\rangle\otimes|N-1\rangle$ ALIS pair. The frequency sweep computed with the essential state model reproduces the AIP splittings from the experiment plot remarkably well. All of the ALIS features observed in the experiment match with energies in the essential states model within 50 meV. 
Fig. ~\ref{fig:avoidedcrossing}c identifies the ALIS features over a broader range of dressing-laser and XUV frequencies. At IR photons energies below 0.8 eV and above 1.1 eV, the bright $|3s^{-1}4p\rangle\otimes|N\rangle$ horizontal feature is traversed by the $|3s^{-1}5p\rangle\otimes|N-2\rangle$ and the $|3s^{-1}4s\rangle\otimes|N+1\rangle$ ALIS, respectively. 

We confirm the assignment of the AIP branches to specific ALIS pairs by selectively eliminating individual resonances from the simulation. \UA{Fig.  \ref{fig:avoidedcrossing}d shows that}, in the absence of the $3s^{-1}3d$ resonance, the AIPs splitting at $\sim$0.9~eV IR energy is greatly reduced, and the AIPs at $\sim$0.8~eV virtually disappear. Since the AIPs around 0.8~eV are only present when both $3s^{-1}3d$ and the $3s^{-1}4f$ resonances are included, we can infer this is due to the two-photon-emission ALIS associated with the $3s^{-1}4f$ state, resonantly enhanced by the $3s^{-1}3d$ intermediate AIS. When the $3s^{-1}5s$ state is removed, the splitting around 0.9~eV is only minimally affected, indicating that most of the contribution in this region comes from the interaction of $|3s^{-1}4p\rangle\otimes|N\rangle$ bright state with the $|3s^{-1}3d\rangle\otimes|N-1\rangle$ ALIS. The $3s^{-1}5s$ state, however, is essential to resonantly enhance the two photon transition responsible for the $|3s^{-1}5p\rangle\otimes|N-2\rangle$ ALIS, giving rise to the avoided crossing around 0.7~eV. 

In conclusion, we have measured the autoionizing polariton multiplets in the delay dependent absorption spectrum of argon, and assigned them with the help of \emph{ab initio} simulations. Experiment and theory are in excellent agreement and indicate the formation of entangled light-matter states with low autoionization rate, due to the destructive interference of the Auger decay amplitude from the different resonant components of the polariton. These results pave the way to improve coherent control protocols in the continuum thanks to laser induced Auger decay stabilization.

LA and CC acknowledge support from the NSF Theoretical AMO grants N~1607588 and N~1912507, and computer time at the University of Central Florida Advanced Research Computing Center. E. L. acknowledges support from the Knut and Alice Wallenberg Foundation, project KAW 2017.0104, and the Swedish Research Council, Grant No. 2016-03789. NH and AS acknowledge support from the U. S. Department of Energy, Office of Science, Office of Basic Energy Science, under award no. DE-SC0018251.

\bibliography{Ar_ATAS_refs}

\end{document}